\documentstyle[12pt]{article}
\def\beq{\begin{equation}}
\def\eeq{\end{equation}}
\def\bea{\begin{eqnarray}}
\def\eea{\end{eqnarray}}

\def\ba{\begin{array}}
\def\ea{\end{array}}
\def\bce{\begin{center}}
\def\ece{\end{center}}

\def\nonu{\nonumber}

\begin{document}
\begin{titlepage}
\rightline{SNUTP-97-115}
\rightline{UM-TG-197}
\rightline{hep-th/9708127}
\def\today{\ifcase\month\or
        January\or February\or March\or April\or May\or June\or
        July\or August\or September\or October\or November\or December\fi,
  \number\year}
\vskip 1cm
\centerline{\Large \bf $Sp(N_c)$ Gauge Theories and M Theory Fivebrane }
\vskip 1cm
\centerline{\sc Changhyun Ahn$^{a,}$\footnote{ chahn@ctp.snu.ac.kr},  
Kyungho Oh$^{b,}$\footnote{ oh@arch.umsl.edu}
and Radu Tatar$^{c,}$\footnote{tatar@phyvax.ir.miami.edu}}
\vskip 1cm
\centerline{{\it $^a$ Dept. of Physics, Seoul National University,
Seoul 151-742, Korea}}
\centerline{{ \it $^b$ Dept. of Mathematics,
 University of Missouri-St. Louis,
 St. Louis, Missouri 63121, USA}}
\centerline{{\it $^c$ Dept. of Physics, University of Miami,
Coral Gables, Florida 33146, USA}}
 \vskip 2cm
\centerline{\sc Abstract}
\vskip 0.2in
We analyze M theory fivebrane in order to study the moduli space of
vacua of $N=1$ supersymmetric $Sp(N_c)$ gauge theories with $N_f$ flavors
in four dimensions.
We show how the $N=2$ Higgs branch can be encoded in M theory by studying
the orientifold which plays a crucial role in our work.
When all the quark masses are the same, the surface of the M theory
spacetime  representing 
a nontrivial ${\bf S^1}$ bundle over ${\bf R^3}$ develops $A_{N_f-1}$ type 
singularities
at two points where D6 branes are located. Furthermore, by turning off the 
masses,
two singular points on the surface collide and produce $A_{2N_f-1}$ type 
singularity.
The sum of the multiplicities of rational curves on the resolved surface 
gives the dimension of
$N=2$ Higgs branch which agrees with the counting from  the 
brane configuration picture of type
IIA string theory.
By rotating M theory fivebranes we get the strongly coupled dynamics of
$N=1$ theory and describe the vacuum expectation values of the meson field 
parameterizing
Higgs branch which are 
in complete agreement with the field theory results. Finally, we take
the limit where the mass of adjoint chiral multiplet goes to infinity and 
compare with
field theory results. For massive case, we comment on some relations with 
recent work
which deals with $N=1$ duality in the context of M theory.    
\end{titlepage}
\newpage
\setcounter{equation}{0}

\section{Introduction}
\setcounter{equation}{0}

One of the most interesting tools used to study nonperturbative 
dynamics of low energy supersymmetric gauge theories is to understand the
D(irichlet) brane dynamics where the gauge theory is realized on the 
worldvolume of
D brane. 

This work was pioneered by Hanany and Witten \cite{hw} where 
the mirror symmetry of $N=4$ gauge theory in 3 dimensions was interpreted 
by changing the position of the Neveu-Schwarz(NS)5 brane in spacetime.
(see also \cite{bo1}\cite{bo2}).
They took a configuration of type IIB string theory 
which preserves 1/4 of the supersymmetry and consists of
parallel NS5 branes with D3 branes suspended between them and D5 branes located
between them. A new aspect of brane dynamics was the creation of D3 brane 
whenever
a D5 brane and NS5 brane are crossing through each other. This was due to
the conservation of the linking number(defined as a total magnetic charge 
for the
gauge field coupled with the worldvolume of the both types of NS and D branes).
 
By T-dualizing the above configuration on one space coordinate, 
the passage to $N=2$ gauge theory in 4 dimensions can be described
as two parallel NS5 branes and D4 branes suspended between them 
in a flat space in type IIA string theory.
When one change the relative orientation of the two NS5 branes \cite{bar} 
while keeping
their common 4 spacetime dimensions intact, the $N=2$ supersymmetry is
broken to $N=1$. The brane configuration\cite{egk,egkrs} 
preserves 1/8 of the supersymmetry
and this corresponds to turning on a mass of adjoint field because the 
distances
between D4 branes suspended between the NS5 branes relate to the vacuum 
expectation values(vevs).
The configuration of D4 branes gives the gauge
group while the D6 branes give the global flavor group.
Using this configuration they described and checked a stringy
derivation of Seiberg's duality for $N=1$ supersymmetric gauge theory with
$SU(N_c)$ gauge group with $N_f$ flavors in the fundamental representation
which was previously conjectured in \cite{se1}.
This result was generalized to brane configurations with orientifolds which 
then
give $N=1$ supersymmetric theories with gauge group $SO(N_c)$ or 
$Sp(N_c)$ \cite{eva,egkrs}.
In this case the NS5 branes have to pass over each other and some strong 
coupling
phenomena have to be considered.
Similar results were obtained in \cite{bh,bsty,t} where the moduli space
of the supersymmetric gauge theories is geometrically encoded in the 
brane setup.

Another approach was initiated by Ooguri and Vafa \cite{ov} where they 
considered the compactification of IIA string theory on a double elliptically
fibered Calabi-Yau threefold.
The wrapped  D6 branes around three cycles of Calabi-Yau threefold filling also
a 4 dimensional spacetime. The transition between electric theory and its 
magnetic
dual appears when a change in the moduli space of Calabi-Yau threefold occurs.
Their results were generalized in the papers of \cite{ao,a,ar,aot} to various 
other models which reproduce
field theory results studied previously. 

So far the branes in string theory were considered to be rigid without any 
bendings.
When the branes are intersecting each other, a singularity occurs. 
In order to avoid that kind of singularities, a very nice
simplification was obtained by reinterpreting  brane configuration in string 
theory
from the point of view of M theory as was showed by Witten in \cite{w1}. 
Then both the D4 branes
and NS5 branes come from the fivebranes of M theory (the former is an M theory
fivebrane wrapped over $\bf{S^1}$ and the latter is an M theory fivebrane on
$\bf{R^{10} \times S^1}$). That is, D4 brane's worldvolume projects to a 
five manifold in
$\bf{R^{10}}$ and NS5 brane's worldvolume is placed at a point in $\bf{S^1}$ 
and
fills a six manifold in $\bf{R^{10}}$. To obtain D6 branes one has to use a 
multiple
Taub-NUT space whose metric is complete and smooth.
The $N=2$ supersymmetry 
in four dimensions requires that the worldvolume of M theory fivebrane is 
$\bf{R^{1,3}}\times \Sigma$ where $\Sigma$ is uniquely identified with the
curves \cite{sw} that appear in the solutions to Coulomb branch of the 
field theory.
The configurations involving orientifolds were considered in\cite{lll,bsty1}.
The method of brane dynamics was used to study supersymmetric field theories 
in several dimensions by many authors 
\cite{ah,ba,k,cvj1,hov,hz,mmm,fs,w2,biksy,gomez,cvj2,hk,hsz,nos,hy,noyy,ss,bo,mi}. 

The original work \cite{w1} was suited to study the moduli space for $N=2$
supersymmetric theories. By rotating one of the NS5 branes the $N=2$ 
supersymmetry is broken to $N=1$ \cite{bar}. 
In \cite{w2,hoo} (see also \cite{biksy,ss,bo}) this was seen from the 
point of the M theory 
interpretation, by considering the possible deformation of the curve $\Sigma$. 
In field theory, the supersymmetry is broken by giving a mass to the
adjoint field and if this mass is finite, the $N=1$ field theory can be 
compared 
with the previous results obtained in \cite{ads}. 
These papers considered the case of unitary groups. 

Recently, the exact low energy description of $N=2$ supersymmetric 
$SU(N_c)$ gauge
theories with $N_f$ flavors in 4 dimensions in the framework of M theory
fivebrane have been found in \cite{hoo}.
They constructed M fivebrane configuration which encodes the information of
Affleck-Dine-Seiberg superpotential \cite{ads} for $N_f < N_c$. Later, 
this approach has been
used to study the moduli space of vacua of confining phase of $N=1$ 
supersymmetric
gauge theories in four dimensions \cite{bo}. In terms of brane 
configuration of IIA string
theory, this corresponds to the picture of \cite{egk} by taking 
multiples of NS'5 branes rather than a single NS'5 brane.

In the present paper we generalize to the case of symplectic group 
$Sp(N_c)$ with
$N_f$ flavors. The new
ingredient that is introduced is the orientifold. We find an interesting
picture which differs from the one obtained for unitary group 
$SU(N_c)$ \cite{hoo}. This is
expected because for $SU(N_c)$ groups we have both baryonic and 
non-baryonic branches, but
in the case of $Sp(N_c)$ we cannot construct any baryon, so we have only 
non-baryonic branch. 

This paper is organized as follows.
In section 2 we review the papers of \cite{ip,aps,hms} and study 
the moduli space of vacua of the $N=1$ theory which is
obtained from the $N=2$ theory by adding a mass term to the adjoint chiral
multiplet. We discuss for different values of the number of flavors 
with respect to the number of colors.  We also introduce massive matter.
In section 3, we start with the setup of M theory fivebrane and 
discuss the Higgs
branches with the resolution of singularities.
In section 4, we rotate brane configuration and obtain information 
about the strong
coupling dynamics of $N=1$ theory.
In section 5, we take the mass of adjoint field infinite and compare 
it with field theory
results for massless or massive matter. We discuss $N=1$ duality and 
compare with
similar work without D6 branes obtained in \cite{cs} recently.
Finally in section 6, we conclude our results and comment on the outlook in the
future directions.

\section{Field Theory Analysis}
\setcounter{equation}{0}

Let us review and summarize field theory results already known in the papers of
\cite{ip,aps,hms} for future developments.
We claim no originality for most of results presented in this 
section except that
we have found the property of meson field $M^{ij}$ having only one 
kind matrix element
which will be discussed in detail later.

\subsection{$N=2$ Theory}
Let us consider $N=2$ supersymmetric  $Sp(N_{c})$ gauge theory
with matter in the $2N_c$ dimensional representation of $Sp(N_c)$. In terms
of $N=1$ superfields, $N=2$ vector multiplet consists of a field strength
chiral multiplet $W_{\alpha}^{ab}$ and a scalar chiral 
multiplet $\Phi_{ab}$, both in the
adjoint representation of the gauge group $Sp(N_c)$. The quark hypermultiplets
are made of a chiral multiplet $Q^{i}_{a}$ which couples to the 
Yang-Mills fields where 
$i = 1,\cdots ,2N_{f}$ are flavor indices( the number of flavors has
to be even ) and $a = 1, \cdots , 2N_{c}$ are color indices. 
The
$N=2$ superpotential takes the form:
\begin{equation}
\label{super}
W = \sqrt{2} Q^{i}_{a} \Phi^{a}_{b} J^{bc} Q^{i}_{c} 
+ \sqrt{2} m_{ij} Q^{i}_{a} J^{ab} Q^{j}_{b},
\end{equation}
where $J_{ab}$ is the symplectic metric 
\bea
 \left( \begin{array}{cc} 
0 & 1 \\ -1 & 0      
\end{array} \right) \otimes {\bf 1_{N_c \times N_c}} 
\eea
used to raise and lower 
$Sp(N_{c})$ color indices and $m_{ij}$ is the antisymmetric mass matrix
\bea
\label{mass}
 \left( \begin{array}{cc}
0 & -1 \\ 1 & 0      
\end{array}  \right) \otimes \mbox{diag} ( m^f_{1}, \cdots, m^f_{N_f} ).
\eea
Classically, the global symmetries are the flavor
symmetry $O(2N_{f}) = SO(2N_{f})\times \bf{Z_{2}}$ in addition to  
$U(1)_{R}\times SU(2)_{R}$ chiral R-symmetry. 
The theory is asymptotically free for $N_{f}$ smaller than $2N_{c}+2$ 
and generates
dynamically a strong coupling scale $\Lambda_{N=2}$ where we denote the 
$N=2$ theory
by writing it in the subscript of $\Lambda$.
The instanton factor
is proportional to $\Lambda_{N=2}^{2N_{c}+2-N_{f}}$. Then the 
$U(1)_{R}$ symmetry is anomalous and is broken down to a discrete 
$\bf{Z_{2N_{c}+2-N_{f}}}$ by instantons.

The moduli space contains the Coulomb and the Higgs branches.
The Coulomb branch is parameterized by the gauge invariant order parameters
\bea
u_{2k}=<\mbox{Tr}(\phi^{2k})>, \;\;\; \mbox{where} \;\; k=1, \cdots, N_c
\eea 
where $\phi$ is the scalar field in $N=2$ vector multiplet.
Up to a gauge transformation $\phi$  can be  diagonalized to a
complex matrix, 
$<\phi>=\mbox{diag} ( A_1, \cdots, A_{N_c} )$ where $A_i=
( { a_i \atop  0 }{  0 \atop  -a_i}  )$.
At a generic point the vevs of 
$\phi$ breaks the $Sp(N_c)$ gauge symmetry
to $U(1)^{N_c}$ and the dynamics of the theory is that of an 
Abelian Coulomb phase. The Wilsonian effective Lagrangian in the low
energy can be made of the multiplets of $A_i$ and $W_i$ where
$i=1, 2, \cdots, N_c$. 
If $k$ $a_i$'s are equal and nonzero then there 
exists an enhanced $SU(k)$ gauge symmetry. When they are also zero, 
an enhanced $Sp(k)$ gauge symmetry appears.
On the other hand, 
the Higgs branches are described by gauge invariant quantities 
which are made from
the squarks vevs and which can be written as the meson field 
$M^{ij} = Q^{i}_{a} J^{ab} Q^{j}_{b}$ because we do not have any baryons.

\subsection{Breaking $N=2$ to $N=1$}

We want now to break $N=2$ supersymmetry down to $N=1$ supersymmetry
by turning on a bare mass $\mu$ for the adjoint chiral multiplet $\Phi$. 
For the moment we consider that all the squarks
are massless, so the terms of $m_{ij}$  in (\ref{super}) will 
not enter into our superpotential. The
superpotential is expressed as follows:
\begin{equation}
\label{e1}
W = \sqrt{2} Q^{i}_{a} \Phi^{a}_{b} J^{bc} Q^{j}_{c} + \mu \; 
\mbox{Tr}(\Phi^{2}).
\end{equation}

When the mass of the adjoint chiral multiplet is much smaller than
$\Lambda_{N=2}$, by turning a mass for
the adjoint chiral multiplet, the structure of moduli space of vacua for 
$N=2$ theory is changed.
Most of the Coulomb branch is lifted except $2N_{c} + 2 - N_{f}$
points which are related to each other by the action of 
${\bf Z_{2N_{c} + 2 - N_{f}}}$ .

When the mass $\mu$ is increased beyond $\Lambda_{N=2}$ we can integrate
out the adjoint chiral multiplet in the low-energy theory. 
Below the
scale $\mu$, by a one loop matching between the $N=1$ and $N=2$ theories
we obtain the $N=1$ dynamical scale, $\Lambda_{N=1}$ to be:
\begin{equation}
\label{scale}
\Lambda_{N=1}^{2(3N_{c}+3-N_{f})} = \mu^{2N_{c} + 2}\Lambda_{N=2}^{2(2N_{c} 
+ 2
-N_{f})}.
\end{equation}

If $\mu$ is much larger than $\Lambda_{N=1}$ but finite, 
we can integrate out the heavy field $\Phi$ and to obtain
a superpotential which is quartic in the squarks and
proportional to $1/\mu$.  
The F-term equation for $\Phi$ from (\ref{e1}) gives us to
\begin{equation}
Q^{i}_{a}Q^{i}_{c} + \sqrt{2}\mu J_{ab}\Phi^{b}_{c} = 0
\end{equation}
where we can read off $\Phi^b_c$
\begin{equation}
\Phi^{b}_{c} = \frac{1}{\sqrt{2}\mu} J^{ab} Q^{i}_{a} Q^{i}_{c}
\end{equation}
or
\begin{equation}
\Phi^{2} = - \frac{1}{2\mu^{2}} M^{2}.
\label{fi}
\end{equation}
We plug this into the superpotential equation and obtain
\begin{equation}
\Delta~W = - \frac{1}{2\mu} \mbox{Tr}(M^{2})
\label{pot}
\end{equation}
which is similar to the equation $(2.5)$ of \cite{hoo} 
but without the term involving $(\mbox{Tr}M)^{2}$ because $M$ is traceless
antisymmetric in our case and with a minus sign due to (\ref{fi}).
Therefore, the system below the energy scale $\mu$ can be regarded as 
the $N=1$ SQCD with the tree level superpotential (\ref{pot}) and with
the dynamical scale $\Lambda_{N=1}$ given by (\ref{scale}). When
we take the limit of $\mu\rightarrow \infty$ keeping $\Lambda_{N=1}$ fixed, 
the superpotential (\ref{pot}) vanishes. 

Let us start with the discussion for the various values of
$N_{f}$ as a function of $N_{c}$.

$\bullet \;\;\; 0\le N_{f}\le N_{c}$

In this range of the number of flavors, as it is well known, a 
superpotential is dynamically generated \cite{ads} by strong coupling
effects.
For a general value of $N_{f}$, the ADS 
superpotential is given by \cite{ip}:
\begin{equation}
\label{ads}
W_{ADS} = (N_{c} + 1 - N_{f}) \; \omega_{N_{c} + 1 - N_{f}}
(\frac{2^{N_{c} - 1}\Lambda_{N=1}^{3(N_{c}+1)-N_{f}}}{\mbox{Pf}M})^{\frac{1}
{N_{c}+1-N_{f}}}
\end{equation}
where $\omega_{N_{c} + 1 - N_{f}}$ is an $N_{c}+1-N_{f}$ th root of unity
and Pf(Pfaffian) of antisymmetric matrix $M$ has the following relation:
$(\mbox{Pf}M)^2=\mbox{det} M$.
 For $N_{f} = N_{c}$, the gauge group is
completely broken for $\mbox{Pf}<M>$ not zero and the ADS superpotential is
generated by an instanton in the broken $Sp(N_{c})$.

For large but finite values for $\mu$, the potential obtained after
$N=2$ breaking into $N=1$ (\ref{pot}) can be described as a perturbation
theory to the ordinary $N=1$ theory. Then the total effective superpotential is
the sum of $W_{ADS}$ and $\Delta ~W$
\beq
W_{eff} = W_{ADS} + \Delta ~W.
\eeq
This form for $W_{eff}$ is exact for any non-zero value of $\mu$.
The argument is based on the holomorphic property. 
The two terms appearing in
an analytic function expanded with respect to $1/\mu$ are given by 
$W_{ADS}$ and
$\Delta$W so a term that can be generated are of the form:
\beq
\label{term}
\mu^{-\alpha}M^{\beta}\Lambda_{N=1}^{(3(N_{c}+1)-N_{f})\gamma}
\eeq
where $\alpha,\gamma$ are non-negative integers. In order to obtain
$\alpha,\beta,\gamma$ we use the fact that (\ref{term}) is invariant
under the axial flavor symmetry $U(1)_{A}$ and has a charge 2 under the
R-symmetry $U(1)_{R}$ where $U(1)_R$ is the anomaly
free combination of the $U(1)$ R-symmetry group. 
The charges for $\Lambda_{N=1}, \mu$ and $M$ are
given by:

\begin{tabular}{||c|c|c|c||} \hline
  & $\Lambda^{3N_{c}+3-N_{f}}_{N=1}$ & M & $\mu$ \\ \hline
 $U(1)_{R}$ & 0 & $\frac{2(N_{f}-N_{c}-1)}{N_{f}}$ & 
 $\frac{2(N_{f}-2N_{c}-2)}{N_{f}}$ \\ \hline    
 $U(1)_{A}$ & $2N_{f}$ & 2 & 4 \\ \hline
\end{tabular}

Using these values for the charges and applying them in (\ref{term}),
the condition that the superpotential has $U(1)_{A}$ charge $0$ gives us that
$2\alpha -\beta=N_{f}\gamma$ and other condition that the superpotential has 
charge $2$ under $U(1)_{R}$ becomes
$N_{f}(-\alpha+\beta-1) = (N_{c}+1)(-2\alpha+\beta)$. The combination of these
two relations leads to:
\beq
1 - \alpha= (N_{c} + 1 - N_{f})\gamma.
\eeq
Because $\gamma\ge 0$ and we are considering the case of $N_{f}<N_{c}$, 
we have only two solutions for this 
equation, that is, $\alpha = 0,\gamma = 1/(N_{c}+1-N_{f})$ and $\alpha=1, 
\gamma=0$,
which exactly correspond to the two terms which appear in (\ref{term}). 
Therefore,
it turns out that the superpotential (\ref{term}) is exact.

The moduli space of vacua is obtained by extrematizing this superpotential,
and we get:
\begin{equation}
\label{moduli}
M^{2} = -\mu \; \omega_{N_{c}+1-N_{f}}(
\frac{2^{N_{c}-1}\Lambda_{N=1}^{3(N_{c}+1)-
N_{f}}}{\mbox{Pf}M})^{\frac{1}{N_{c}+1-N_{f}}}.
\end{equation}

In this moment, by a similarity transformation, $M$ can be brought to different
forms. In \cite{aps,hms}, $M$ has been brought to a form such that
$M^{2}=0$ which is the right equation for $M$ whenever  we do not
consider the ADS potential. But in our case, the equation (\ref{moduli}) tells
us that $M^{2}$ is not equal with $0$, so we take another form for $M$
after a similarity transformation. We bring $M$ to the simplest form,
i.e., with two top-right and bottom-left diagonal blocks, one being 
minus the other because
$M$ is to be antisymmetric. Denote $m_{1}, \cdots,  m_{N_{f}}$ by  the 
elements of
top-right diagonal block in $M$ and of course $-m_{1}, \cdots, -m_{N_{f}}$ 
by the
elements of the bottom-left diagonal block. In this case we will have an 
equation
like (\ref{moduli}) for each $m_{i}$. Since the right hand side is the same for
all the $m_{i}$'s, {\it  they have to be equal}. So all the
diagonal entries in the top-right and bottom-left diagonal blocks of {M} 
are equal. 
This is our {\it new} observation 
which will appear naturally in section 4 due to the symmetry
of orientifolding
and can be compared with the result of \cite{hoo} for $SU(N_c)$ case where
there were two cases, one with equal diagonal entries and th other  with
two different entries on the diagonal. 

Now since all the top-right diagonal entries are equal with $m \equiv m_1=
\cdots = m_{N_f}$, 
we find the value for $m$ by solving (\ref{moduli}):
\begin{equation}
\label{value}
m = 2^{\frac{N_{c}-1}{2(N_{c}+1)-N_{f}}}
\mu\Lambda_{N=2}  
\end{equation}
where we have used the renormalization group( RG ) matching 
equation (\ref{scale}). 
The values of $m$ in equation (\ref{value}) describe the moduli space of the
$N=1$ theory in the presence of a perturbation to the ADS superpotential.
When $\mu\rightarrow\infty$ and $\Lambda_{N=1}$ are finite, the solution
diverges. In this case $\Delta$W is $0$ and divergence of the solution
coincides with the fact that there is no supersymmetric vacua in 
this region of the flavor.   

Let us turn on quark mass terms like
$\frac{1}{2}m^{ij}M_{ij}$. In this case the effective superpotential 
is given by
\beq
W_{eff} = W_{ADS} + \Delta~W + \frac{1}{2}m~M
\eeq
where $W_{ADS}$ and $\Delta ~W$ are given by (\ref{ads}) and (\ref{pot}) and
m is an antisymmetric matrix as in (ref{mass}) but where we take 
$m_{f}=m_{1}^{f} = \cdots = m_{N_{f}}^{f}$
In this case the  equation (\ref{moduli}) is modified to contain a term 
$ \mu m M/2$. 
As we consider the limit $\mu\rightarrow\infty$ by keeping $\Lambda_{N=1}$
finite, the system will be $N=1$ SQCD with massive flavors. Only terms 
which are proportional to $\mu$ will resist (so the term form the LHS
of (\ref{moduli}) will be neglected because it does not depend on $\mu$)
and thus we obtain the solution for the moduli space to be:
\beq
\omega_{N_{c}+1-N_{f}}
(\frac{2^{N_{c}-1}\Lambda_{N=1}^{3(N_{c}+1)-N_{f}}}{
\mbox{Pf}M})^{\frac{1}{N_{c}+1-N_{f}}}
=  m_{f} m/2
\eeq
with the solution
\beq
m^{N_{c}+1}=\frac{2^{2N_{c}-N_{f}} \Lambda_{N=1}^{
3(N_{c}+1)-N_f}}{m_{f}^{N_{c}+1-N_{f}}}
\eeq
giving $N_{c}+1$ vacua in accordance with the interpretation of the low
energy physics as the pure $N=1$ Yang-Mills theory.

Next we  are now increasing the number of flavors. 

$\bullet \;\;\; N_{f} = N_{c}+1$. 


Now it is obvious that the ADS superpotential vanishes  and the classical
moduli space of vacua is changed quantum mechanically. It is 
parameterized by the meson satisfying the constraint\cite{ip}
\begin{equation}
\mbox{Pf}M = 2^{N_{c}-1} \Lambda_{N=1}^{2(N_{c}+1)}.
\label{cons}
\end{equation} 
Again the quartic term $\Delta ~W$ is small for large finite $\mu$ and can
be considered as a perturbation to the ordinary $N=1$ theory. By introducing
a Lagrange multiplier $X$ in order to impose the constraint(\ref{cons}),
the effective superpotential will be:
\begin{equation}
\label{spt1}
W_{eff} = X (\mbox{Pf}M - 2^{N_{c}-1}\Lambda_{N=1}^{2(N_{c}+1)}) - 
\frac{1}{2\mu}
\mbox{Tr}(M^{2}).
\end{equation}
>From the derivative with respect to $M$, we get:
\begin{equation}
M^{2} = \mu \; X \mbox{Pf}M.
\end{equation}
For the case of $X \neq 0$, again we can bring $M$ by a similarity 
transformation to
the same form as before and this tells us that again all the top-right 
diagonal 
entries are the same and we obtain:
\begin{equation}
m^{N_{c}+1} = 2^{N_{c} - 1}\Lambda_{N=1}^{2(N_{c}+1)}
\end{equation}
which leads to, after using the RG equation:
\begin{equation}
m = 2^{\frac{N_{c}-1}{N_{c}+1}}\mu\Lambda_{N=2}
\end{equation} 
which gives the moduli space.

$\bullet \;\;\; N_{f}=N_{c}+2$

In this case, the effective potential by adding $\Delta ~W$ is given by:
\begin{equation}
W_{eff} = -\frac{\mbox{Pf}M}{2^{N_{c}-1}\Lambda_{N=1}^{2N_{c}+1}}-
\frac{1}{2\mu} \mbox{Tr}(M^{2})
\end{equation}
which give us after extrematizing:
\begin{equation}
M^{2} = -\mu \frac{\mbox{Pf}M}{2^{N_{c}-1}\Lambda_{N=1}^{2N_{c}+1}}.
\end{equation}
Again $M$ can be brought to a simple form by a similarity transformation
and all the diagonal entries are equal. After using the
RG equation we get the moduli space given by:
\begin{equation}
m = 2^{\frac{N_c-1}{N_c}} \mu\Lambda_{N=2}.
\end{equation}

$\bullet \;\;\; N_{f} > N_{c}+2$

The theory that we have discussed until now
is the electric theory which for this range of the number of flavors 
has a dual description in terms of a
$Sp(N_{f}-N_{c}-2)$ gauge theory with $N_{f}$ flavors $q^{i}$ in the
fundamental $(i = 1, \cdots, 2N_{f})$, gauge singlets $M_{ij}$ and a
superpotential
\begin{equation}
W = \frac{1}{4\lambda}M_{ij}q^{i}_{c}q^{j}_{d} J^{cd}.
\end{equation}
where the scale $\lambda$ relates the scale $\Lambda_{N=1}$ of the electric
theory and the scale $\tilde{\Lambda}_{N=1}$ of the magnetic theory by:
\begin{equation}
\Lambda_{N=1}^{3(N_{c}+1)-N_{f}}\tilde{\Lambda}_{N=1}^{3(N_{f}-N_{c}-1)-N_{f}}
= C (-1)^{N_{f}-N_{c}-1} \lambda^{N_{f}}
\end{equation}
where the constant $C$ was found in \cite{ip} to be $C=16$.
The effective superpotential is given as:
\beq
W_{eff} = W + W_{ADS} + \Delta ~W.
\eeq
If the vevs for the magnetic quarks are $0$, then the analysis is identical to 
those for the case $N_{f} < N_{c}$. If the vevs are not zero, then
as in \cite{hoo} we can take a limit to approach $\mbox{Pf}M=0$ and to use 
the corresponding formula in order to compare with the M theory approach.
In the $SU(N_c)$ case, where baryon exist,  a specific choice has 
been taken such that the baryons have a specific interpretation in the 
M theory picture. 

\section{$N=2$ Higgs Branch from M Theory }
\setcounter{equation}{0}

In this section we study the moduli space of vacua of $N=2$ supersymmetric
QCD by analyzing M theory fivebranes. We will consider the Higgs branch 
in terms of
geometrical picture. Let us first describe the Higgs branch in the type
IIA brane configuration.

Following the paper of \cite{egk}, the brane
configuration contains three kind of branes: the two parallel NS5
branes extend in the direction $(x^0, x^1, x^2, x^3, x^4, x^5)$, the D4 branes
are stretched between two NS5 branes and extend over $(x^0, x^1, x^2, x^3)$ and
are finite in the direction of $x^6$, and the D6 branes extend in the direction
of $(x^0, x^1, x^2, x^3, x^7, x^8, x^9)$. In order to study symplectic or
orthogonal gauge groups, we will consider an O4
orientifold which is parallel to the D4 branes in order to keep the
supersymmetry and is not of finite extent in $x^6$ direction. The D4 branes
is the only brane which is not intersected by this O4 orientifold. 
The orientifold
gives a spacetime reflection as $(x^4, x^5, x^7, x^8, x^9) \rightarrow
(-x^4, -x^5, -x^7, -x^8, -x^9)$, in addition to the gauging of worldsheet
parity $\Omega$. The fixed points of the spacetime symmetry define 
this O4 planes.
Each object which does not lie at the fixed points ( i.e. over the orientifold
plane), must have its mirror image. Thus NS5 branes have a mirror in $(x^4,
x^5)$ directions and D6 branes have a mirror in $(x^7, x^8, x^9)$ directions.
Another important aspect of the orientifold is its charge, given by the charge
of $H^{(6)}=d A^{(5)}$ coming from Ramond Ramond(RR) sector, which is
related to the sign of $\Omega^2$. In the natural normalization, where the
D4 brane carries one unit of this charge, the charge of the O4 plane is $
\pm 1$, for $\Omega^2= \mp 1$ in the D4  brane sector.

With the above preliminary setup, let us discuss about the two different 
branches of the theory. The Coulomb
branch can be described when all the D4 branes lie between NS5 branes where no
squark has vevs. To go to the Higgs branch, the D4 branes
are broken on the D6 branes and are suspended between D6 branes
being allowed to move on the
$(x^{7}, x^{8}, x^{9})$ directions. Together with the gauge field component
$A_{6}$ in the $x^{6}$ coordinate this gives two complex parameters to
parameterize the location of the D4 branes. 
In \cite{hoo}, for $SU(N_{c})$ case, the Coulomb branch and
the Higgs branch share common directions and this comes from the fact that
there are two different eigenvalues for $M$ which correspond to 
$r$ equal eigenvalues and
$N_{c}-r$ equal eigenvalues. 
By turning on vevs for $r$ squarks, this gives rise to
make the $r$ dimensional block of $M$ be nonzero. In brane language, this 
describes breaking $r$ D4 branes on the D6 branes and suspending the remaining 
$N_{c}-r$ D4 branes
between the two NS5 branes.

In the case of $Sp(N_c)$ gauge theory, there are only 
two possibilities:

${\bullet}$ All D4 branes are suspended between the two NS5 branes where no
squark has vevs.

${\bullet}$ Some of D4 branes are broken on D6 branes \footnote{
There are recent papers on this issue \cite{extra,extra1}.}.
 
We, for simplicity, restrict ourselves to the case of
all D4 branes being broken on D6 branes. See \cite{extra,extra1} for more
general cases.
 
The motion of D4 branes along 
D6 branes describes the Higgs branch and for each D4 brane suspended between
two D6 branes there exist two massless complex scalars parameterizing the
fluctuations of the D4 brane. Because of the O4 orientifold we have
to take into account D4 branes stretched between two D6 branes.
 The s-rule \cite{hw} allows only one D4 brane (and its
mirror) between a NS5 brane and a single D6 brane (and its mirror).
For $N=2$ theory because we have two NS5 branes, for both of them we have
to impose the s-rule. Also, in contrast with $N=1$ theory, 
there are no complex scalars which correspond to D4 branes stretched
between NS'5 branes and D6 branes. However remember that
for $N=1$ theory\cite{egkrs}  there are no complex
scalar corresponding to a D4 brane stretched between NS5 brane and a D6 brane.
The dimension of the Higgs moduli space is obtained 
by counting all possible breakings of D4 branes on D6 branes 
as follows: the first D4 brane is broken in
$N_{f}-1$ sectors between the D6 branes (therefore the complex dimension 
is the 
twice of $N_f-1$), the second D4 branes is
broken in $N_{f}-3$ sectors (the complex dimension is twice of $N_f-3$) 
and so on. But, besides that we have to consider the antisymmetric orientifold
projection which eliminates some degrees of freedom, as explained in
\cite{egkrs}. 
Then the dimension of the Higgs moduli space is given by:
\beq
\label{higgsdim}
2[(2N_{f}-2-1) + (2N_{f}-6-1) + \cdots + (2N_{f}-4N_{c}+2-1)]=
4N_{c}(N_{f}-N_{c})-2N_{c}
\eeq
or $4N_{c}N_{f} - 2N_{c}(2N_{c}+1)$ where in the previous equations we have
explicitly extracted 1 as a result of the antisymmetric orientifold 
projection.The overall factor $2$ in the left hand side is due 
to the mirror D6 branes and the result is 
very similar to the field theory result except 
an extra multiplicative factor 2 in the right hand side, because we consider
here complex dimensions.
In field theory, because of the $N_{f}$ vevs, 
the gauge symmetry is completely broken and there are 
$4N_{c}N_{f}-2N_{c}(2N_{c}+1)$ massless neutral hypermultiplets 
for a $N=2$ supersymmetric theory which thus exactly gives the
dimension of the Higgs moduli space. Thus, the field theory results match
the brane configuration results.

Let us discuss how the above brane configuration appears in M theory 
context in terms of
a generically smooth single M fivebrane whose worldvolume is
${\bf R^{1,3}} \times \Sigma$ where $\Sigma$ is identified with Seiberg-Witten
curves\cite{as} that determine  the solutions to Coulomb branch of 
the field theory.
As usual, we write $v=x^4+i x^5, s=(x^6+i x^{10})/R, t=e^{-s}$
where $x^{10}$ is the eleventh coordinate of M theory which is compactified
on a circle of radius $R$. Then the curve $\Sigma$, describing
 $N=2$  $Sp(N_c)$ gauge theory with $N_f$ flavors,
is given\cite{lll} by an equation in $(v, t)$ space
\bea
\label{ah0}
t^2-(v^2 B(v^2, u_k)+  \Lambda_{N=2}^{2N_c+2-N_f } \prod_{i=1}^{N_f} m_i )t+
\Lambda_{N=2}^{4N_c+4-2N_f} \prod_{i=1}^{N_f} (v^2-{m_i}^2)=0
\eea
where $B(v^2)$ is a polynomial of $v^2$ of degree $N_c$ with the coefficients
depending on the moduli $u_k$, $
v^{2N_c} + u_2 v^{2N_c -2} + \cdots + u_{2N_c}
$ and $m_i$ is the mass of quark\footnote{Note that this $m_i$ is nothing to do with the
element of meson field $M$. Unfortunately we used same notation.}.

\subsection{Including D6 Branes }

In M theory, the type IIA D6 branes are the magnetic dual of the 
electrically charged  D0 branes, 
which  are the Kaluza-Klein monopoles described by a Taub-NUT space. 
This is derived
from a hyper-K\"ahler solution of the four-dimensional Einstein equation. 
But we will ignore the
hyper-K\"ahler structure of this Taub-NUT space. Instead, we use one  of the 
complex structures, which
 can be  described by\cite{lll} 
\bea
\label{ah1}
y z=\Lambda_{N=2}^{4N_c+4-2N_f} \prod_{i=1}^{N_f} (v^2-{m_i}^2)
\eea
in $\bf C^3$. The D6 branes are located at $y=z=0, v=\pm m_i$.
This surface, which represents a nontrivial $\bf S^1$ bundle over 
$\bf R^3$ instead of
the flat four dimensional space ${\bf R^3} \times {\bf{S^1}}$ with coordinates
 $(x^4, x^5, x^6, x^{10})$,
  is the unfolding of the $A_{2n-1}$ ($n=N_f$) singularity in general. 
 The Riemann surface $\Sigma$ is embedded as a curve in this 
curved surface and
given by
\bea
\label{ah2}
y+z=v^2 B(v^2)+  \Lambda_{N=2}^{2N_c+2-N_f } \prod_{i=1}^{N_f} m_i.
\eea
which reproduces to eq. (\ref{ah0}) as we identify $y$ with $t$.
>From the symmetries existent in the type II A brane configuration, not all
of them are preserved in the M-theory configuration. Our type IIA brane
configuration has $U(1)_{4,5}$ and $SU(2)_{7,8,9}$ symmetries interpreted as
classical $U(1)$ ans $SU(2)$ R-symmetry groups of the 4 dimensional theory
on the brane worldvolume. The classical brane configuration is invariant both
under the rotations. One of them, 
only $SU(2)_{7,8,9}$ is preserved in M theory quantum mechanical 
configuration but
$U(1)_{4,5}$ is broken. This is a the same as saying that the $U(1)_{R}$ 
symmetry of the $N=2$ supersymmetric field theory is anomalous being broken by
instantons. As discussed in section 2, the instanton factor is proportional 
with $\Lambda_{N=2}^{2N_{c}+2-N_{f}}$. So we have to see what is the charge
of this factor under $U(1)_{4,5}$. We see this from equations (\ref{ah1}) and
(\ref{ah2}) by considering $v$ of charge 2. We list below the 
charges of coordinates
and parameter in the table:

\begin{tabular}{||c|c|c|c||} \hline
$z$ &$ y$ &$ v$ & $\Lambda_{N=2}^{2N_{c}+2-N_{f}}$ \\ \hline
$4N_{c}+4$ & $4N_{c}+4$ & 2 & $4N_{c}+4-2N_{f}$ \\ \hline
\end{tabular}

In this case, the full $U(1)_{4,5}$ symmetry is restored, by assigning 
the instanton charge $(4N_{C}+4-2N_{f})$ to the $\Lambda$ factor.

Note that whenever some $m_i$ are the same, the smooth complex 
surface (\ref{ah1}) develops
A-type singularity. But this is misleading since the hyper-K\"ahler 
structure becomes 
singular only if the
D6 branes have the same position in $x^6$ and not only in $v$. 
When D6 branes with the 
coincident $m_i$'s  are separated in
the $x^6$ direction, the singular surface (\ref{ah1}) must be replaced 
by a smooth one 
which is the
resolution of $A$-type singularity. We will briefly describe the 
resolution of the 
$A$-type singularity.
On the resolved surface, we also describe the parity due to orientifolding.

\subsection{Resolution of the $A$-type Singularity}

When all bare masses are the same but not zero (say $ m=m_i$), 
the surface  (\ref{ah1}) 
$S$ develops
 singularities of type $A_{n-1}$ 
at two points $y=z=0, v=\pm m$. By succession of blowing ups, 
we obtain a smooth 
complex surface $\tilde{S}$
which 
isomorphically maps onto the singular surface $S$ except at the inverse 
image of the 
singular points. 
Over each singular point, there exist $n-1$ rational curves $\bf CP^1$'s 
on the smooth 
surface $\tilde{S}$. 
These rational curves are called the exceptional curves.
Let us denote the exceptional curves over the point $y=z=0, v=m$ by $C_1, C_2, 
\cdots , C_{n-1}$ and those over
the point $y=z=0, v=-m$ by $C'_1, C'_2, \cdots , C'_{n-1}$. 
Here $C_i$'s (resp. $C'_i$)
 are arranged so that $C_i$ (resp. $C'_i$) intersects $C_j$ (resp. $C'_j$)
only if $i= j\pm 1$. The symmetry due to orientifolding yields the 
correspondence 
between $C_i$ and $C'_i$.

When we turn off the bare mass, that is, $m_i =0$ for all $i$, the 
singularity gets worse.  
Two singular points on
the surface $S$
collides to create the $A_{2n-1}$ singularity.  Now there are ${2n-1}$ 
exceptional curves 
on the resolved
surface, which may be considered as a union of two previous 
exceptional curves $C_i$ 
and $C'_i$
and a new rational curve, say $E$ which connects these two exceptional curves.
The orientifold provides
a reflection between $C_i$ and $C'_i$ while inducing a self-automorphism 
on $E$. 
The more precise picture of the resolved surface is as follows: It is
 covered by $2N_f$ complex planes $U_1, U_2, \cdots , U_{2n}$
with coordinates $(y_1 =y, z_1), (y_2, z_2) ,\cdots , (y_n, x_n =y)$ 
which are mapped
to the singular surface $S$ by
\bea
U_i \ni (y_i, z_i) \mapsto  \left\{ \begin{array}{l}
		y= y^i_i z^{i-1}_i\\
		z = y_i^{2N_f-i}z_i^{2N_f+1-i}\\
		v= y_iz_i
			\end{array} \right.
\eea
The planes $U_i$ are glued together by $z_iy_{i+1} =1$ and $y_iz_i = 
y_{i+1}z_{i+1}$. 
The exceptional curve $C_i$ is defined by the locus of 
$y_i=0$ in $U_i$ and $z_{i+1}=0$ in $U_{i+1}$, 
the exceptional  curve $E$ by $y_n=0$ in $U_n$ and $z_{n+1} =0$ in 
$U_{n+1}$ and the
exceptional  curve $C'_i$ by
$y_{2n-i}=0$ in $U_{2n-i}$ and $z_{2n-i+1}=0$ in $U_{2n-i+1}$.
The separation
of the D6 branes in the $x^6$ direction corresponds to the infinitesimal 
direction on the 
singular surface $S$. 
Hence the position of the $D6$-brane may be interpreted as the $2N_f$ 
intersection points 
of the exceptional curves.

\subsection{The Higgs Branch}
In this section, all the bare masses are turned off.
In M theory, the transition to the Higgs branch occurs when the 
fivebrane intersects
 with the $D6$-branes,
which means that the curve $\Sigma$ given by (\ref{ah2}) passes 
through the singular 
point $y=z=v=0$. As a special case, we will consider
  when all D4 branes are broken on $D6$ branes in type IIA picture.
Write  the right hand side of (\ref{ah2}) as:
\bea
\label{C}
v^2B(v^2) = v^2(v^{2N_c} + u_2 v^{2N_c -2} + \cdots + u_{2N_c}).
\eea
Then  our case corresponds to  $u_k =0$ for all $k$. 
To describe the Higgs branch, we will study 
how the curve 
\bea
y+z = v^{2N_c}
\eea
looks like in the resolved $A_{2N_f -1}$ surface. Here we ignored the 
factor $v^2$ in 
the right hand side of
(\ref{C}) because it is always contained in the orientifold 
plane $O4$ and thus does not
contribute to the Higgs branch.
Away from the singular point $y=z=v=0$, we may regard the curve 
as embedded in the 
original
 $y-z-v$ space because there is no change in the resolved surface 
in this region.
Near the singular point $y=z=v=0$, we have to consider the resolved surface.  
On the $i$-th patch  $U_i$ of the resolved
surface, the equation  of the curve $\Sigma$ becomes
\bea
y^i_i z^{i-1}_i +  y_i^{2N_f-i}z_i^{2N_f+1-i} =  y_i^{2N_c}z_i^{2N_c}
\eea
Now  we may factorize this equation according to the range of $i$:
For $i=1,\ldots , 2N_c$, we have
\bea
 y^i_i z^{i-1}_i (1 + y_i^{2N_f -2i}z_i^{2N_f+ 2-2i} - y_i^{2N_c-i} 
z_i^{2N_c+1-i} ) =0 ,
\eea
for $ i=2N_c +1, \ldots , 2N_f -2N_c$,
\bea
y_i^{2N_c} z_i^{2N_c} (y_i^{i-2N_c } z_i^{i-2N_c-1} + 
y_i^{2N_f -i -2N_c}z_i^{2N_f -i-2N_c +1} - 1) = 0 ,
\eea
and for $ i=2 N_f- 2N_c +1,\ldots  ,2N_f$,
\bea
y_i^{2N_f-i}z_i^{2N_f+1-i} ( y_i^{2i-2N_f} z_i^{2i-2 -2N_f} + 
1 - y_i^{2N_c-2N_f +i}z_i^{2N_c-2N_f +i -1}) =0.
\eea
Thus the curve consists of several components. One component, 
which we call $C$, is the zero of the
last factor of the above equations. This extends to the one in 
the region away from 
$y=z=v=0$ which
we have already considered. The other components are the rational 
curves $C_1, 
\ldots , C_{n-1}, E,
C'_1,\ldots , C'_{n-1}$ with some multiplicities. For convenience, we 
rename the exceptional
curves $E_1, \ldots , E_{2n-1}$ so that $E_i$ is defined by $y_i=0$ on 
$U_i$ and 
$z_{i+1} =0$ on $U_{i+1}$.
Hence we can see from the above factorization that the component $E_i$ has  
multiplicity $l_i = i$ for 
$i=1,\ldots , 2N_c $; $l_i =2N_c $ for $i=2N_c +1, \ldots , 2N_f -2N_c $; and 
$l_i = {2N_f-i}$ for
$i=2 N_f- 2N_c +1,\ldots  2N_f-1$.   
 Note that the component $C$ intersects with $E_{2N_c}$ and $E_{2N_f -2N_c}$.

To count the dimension of the Higgs branch, recall
that once the curve degenerates and $\bf CP^1$ components are generated, 
they can move
in the $x^7, x^8, x^9$ directions \cite{w1}.
 This motion together with the integration of the chiral two-forms on such
$\bf CP^1$'s parameterizes the Higgs branch of the four-dimensional theory.
However, we have to omit the component $E_{N_f} =E$ because this corresponds to
the D4 brane connecting a D6 brane and its mirror. (Recall that
$E$ was created after collision of two singular points
 which were mirror to each other.)
 Such a D4 brane is eliminated
by the antisymmetric orientifold projection. Hence we have to put
$l_{N_f} =0$. 
Now, after consideration of
$\bf Z_2$ symmetry, the quaternionic dimension of the Higgs branch is 
\bea
\frac{1}{2}\sum_{i=1}^{2N_f-1} l_i= \sum_{i=1}^{2N_c} i + (2N_f-4N_c )N_c=
2N_c(N_f-N_c)  -N_c,
\eea
which is the half of the complex dimension given in (\ref{higgsdim}). 
 Perhaps, a more appropriate
geometric setting would have been a double covering of a $A_{N_f-1}$ 
singular surface
with the
embedded Seiberg-Witten curve. We will leave this for future investigation.

\section{The Rotated Configuration}
\setcounter{equation}{0}

As we have seen that $N=2$ supersymmetry can be broken to $N=1$ by inserting
a mass term of the adjoint chiral multiplet in field theory approach, 
we analyze
the corresponding configuration in M theory fivebranes. While in the context of
IIA picture, this turns out to be the rotation of one of NS5 branes,
in order to describe this configuration, let us introduce a complex coordinate
\bea
w=x^8+i x^9.
\eea
Of course, the fivebranes are positioned at $w=0$ before the rotation.
Notice that the D4 brane corresponds classically to an M fivebrane at 
$v=w=0$ and
is extended in the direction of $s$, the NS5 brane is at $s=w=0$
and extended in $v$ and NS'5 brane is at $v=0, s=s_0$ and extended in $w$.
Now we rotate only the left NS5 branes and from the behavior of
two asymptotic regions which correspond to the two NS5 branes with $v 
\rightarrow 
\infty$ this rotation leads to the following boundary conditions.
\bea
\label{bdy-cond}
& & w \rightarrow \mu v \;\;\; \mbox{as}\;\;\; v \rightarrow \infty, 
\;\;\; t \sim v^{
2N_c+2}  \nonu \\
& & w \rightarrow 0 \;\;\; \mbox{as}\;\;\; v \rightarrow \infty, \;\;\; t \sim
\Lambda_{N=2}^{2(2N_c+2-N_f)}v^{2N_f-2N_c-2}
\eea
where the left(right) NS5 brane is related to the first(second) 
asymptotic boundary condition.
Far from the origin of the $(v, w)$ plane which is the location of D4 branes,
the location of the NS5 brane in the $s$ plane can be described by
$s(NS5)=-2 R(N_f-N_c-1) \ln v$ while the NS'5 brane in the $s$ plane by
$ s(NS'5)=-2 R(N_c+1) \ln w$.
We discuss about the R-symmetries of the rotated configuration. After
rotation, $SU(2)_{7,8,9}$ is broken to $U(1)_{8,9}$.
In order this to be true, because of the connection between $v$ and $w$ in
(\ref{bdy-cond}), $\mu$ 
has to have charges under $U(1)_{4,5}\times U(1)_{8,9}$. $v$ has charge 2
under $U(1)_{4,5}$ while 0 under $U(1)_{8,9}$. $w$ has charge 0 
under $U(1)_{4,5}$
and 2 under $U(1)_{8,9}$. So $\mu$ has $(-2,2)$ charges under 
$U(1)_{4,5} \times U(1)_{8,9}$. From the equations (\ref{ah1}), (\ref{ah2}) and
(\ref{bdy-cond}) we find the following values for the R-symmetry charges:

\begin{tabular}{||c|c|c|c|c|c|c||} \hline
               & $v$           & $w$ & $y=t$ & $z$               
& $\mu $            
     & $\Lambda_{N=2}^{2N_{c}+2-N_{f}}  $
\\ \hline
$U(1)_{4,5}$ & 2 & 0    & $4(N_{c}+1)$ & $4(N_{c}+1)$ & -2 & 
$4(N_{c}+1)-2N_{f}$ \\ \hline
$U(1)_{8,9}$ & 0 & 2    & 0                & 0                & 2 & 0 \\ \hline
\end{tabular}

Since the rotation is only possible at points in moduli space at which all 
1-cycles 
on the curve $\Sigma$
are degenerate \cite{sw}, the curve $\Sigma$ is rational, which means that the
 functions $v$ and $t$ can
be expressed as a rational functions of $w$ after we identify $\Sigma$ with a 
complex plane $w$ with
some deleted points. 
Because of the symmetry of $v \rightarrow -v, w \rightarrow -w$ due to 
orientifolding, we can write:
\bea
v^2=P(w^2), \;\;\;\;\; t=Q(w^2).
\eea
Since $v$ and $t$ become infinity only if $w=0, \infty$, these rational 
functions are 
polynomials of $w$ up to
a factor of some power of $w$: $P(w^2) = w^{2a}p(w^2), Q(w^2) = 
w^{2b}q(w^2)$ where 
$a$ and $b$ are some 
integers and $p(w^2)$ and $q(w^2)$ are some polynomials of $w$ with 
only even degree 
terms which we may assume 
non-vanishing at $w=0$. Near one of the points at $w=\infty$, $v$ and 
$t$ behave 
as $v\sim \mu^{-1}w$ and
$t \sim t^{2N_c +2}$ by (\ref{bdy-cond}). Thus the rational functions 
are of the form
\bea
P(w^2) = w^{2a}(w^{2-2a} + \cdots)/\mu^{2}\;\;\; \mbox{and}\;\;\; 
Q(w^2) = \mu^{-2N_c -2}w^{2b}(w^{2N_c +2-2b} + \cdots).
\eea
Around a neighborhood $w=0$, the Riemann surface $\Sigma$ can be parameterized 
by $1/v$ which goes to zero as 
$w\to 0$.
Since $w$ and $1/v$ are two coordinates around the neighborhood $w=0$ in the 
compactification of $\Sigma$
and vanish at the same point, they must be linearly related 
$w \sim 1/v$ in the 
limit $w \to 0$.
The function $P(w^2)$ then takes the form
\bea
P(w^2) = \frac{(w^2 -w_+)(w^2 - w_-)}{\mu^2 w^2}.
\eea
However the equation $v^2 = P(w^2)$ implies that $P(w^2)$ must be a square.
Hence we have $w_+ = w_-$  and by letting $w_0^2 =w_{\pm}$
\bea
\label{rot2}
P(w^2)=\frac{(w^2-w_0^2)^2}{\mu^2 w^2}
\eea
which is a square of $w^2-w_0^2/\mu w$.
Since $t\sim  v^{2N_f -2N_c -2}$ and $w\sim 1/v$ as $w\to 0$, we get 
$b= N_c +1 -N_f$ and thus,
\bea
Q(w^2) = \mu^{-2N_c -2}w^{2(N_c+1 -N_f)}(w^{2N_f} + \cdots ).
\eea
For $N_f >0$, by the equation $yz = v^{2N_f}$ defining  the space-time, 
$t=0$ (i.e. $y=0$) implies
$v=0$. Therefore the zeros of the polynomial $w^{2n_f} + \cdots$ are 
$\pm w_0$ of $P(w^2)$. Hence we have
\bea
Q(w^2)=\mu^{-2N_c-2} w^{2(N_c+1-N_f)} (w^2-w_0^2)^{N_f}
\eea
The value of $w_0$ can be determined by the fact that $v^2$ and $t$ 
satisfy the relation
\bea
t+\Lambda_{N=2}^{4N_c+4-2N_f}
v^{2N_f}/ t=v^2 B(v^2)+  \Lambda_{N=2}^{2N_c+2-N_f } \prod_{i=1}^{N_f} m_i
\eea
Then by plugging $v^2$ and $t$ into the above equation
we can read off $w_0$ from the lowest order term  in power of $w$
\bea
&& \mu ^{-2N_c-2} w^{2(N_c+1-N_f)} (w^2-w_0^2)^{N_f}+
\Lambda_{N=2}^{4N_c+4-2N_f}
v^{2N_f} \mu ^{2N_c2} w^{-2(N_c+1-N_f)} (w^2-w_0^2)^{-N_f}= \nonu \\
&& -\frac{(w^2-w_0^2)^2}{\mu^2 w^2} B(v^2)-  
\Lambda_{N=2}^{2N_c+2-N_f } \prod_{i=1}^{N_f} m_i
\eea

We want to calculate now $w_{0}$ from the above equation. 
For this, we will match the lowest order term in powers of $w$ in this 
equation.
Actually we will look for terms with $w^{0}$ i.e., constant terms.
In the left hand side the first term will always have a power of $w$, 
so does not contribute
to the lowest order term. In the right hand side the last term will contain 
at least $w^{2(N_{c}-1)}$ so again does not contribute.
After using the expression for $v$, the second term in left hand side will be
\begin{equation}
\label{w1}
\Lambda_{N=2}^{4N_c+4-2N_f}(w^{2}-w_{0}^{2})^{N_{f}}
\mu^{2N_{c}+2-2N_{f}}w^{-2N_{c}-2}
\end{equation}
In the right hand side, 
from the explicit form of $B(v^{2})$, the only term that can be 
independent of $w$
is obtained when we take only the highest power of $v$  which then will give
$v^{2N_{c}}$. The contribution of this in the right hand side is as follows:
\begin{equation}
\label{w2}
\frac{(w^{2}-w_{0}^{2})^{2N_{c}+2}}{\mu^{2N_{c}+2}}w^{-2N_{c}-2}
\end{equation}
We now extract the lowest order from (\ref{w1}) and (\ref{w2}) and make
them equal to obtain the relation for $w_{0}$ finally:
\begin{equation}
\label{w3}
(-1)^{N_{f}} \Lambda_{N=2}^{4N_c+4-2N_f}
w_{0}^{2N_{f}}\mu^{4N_{c}+4-2N_{f}} = w_{0}^{4N_{c}+4}
\end{equation}
This gives us the value for $w$ to be
\begin{equation}
\label{w4}
w_{0} = (-1)^{\frac{N_{f}}{4N_{c}+4-2N_{f}}}(\mu\Lambda_{N=2}).
\end{equation}
This gives us 
$w_{0}$ up to a $\bf{Z_{4N_{c}+4-2N_{f}}}$ rotation. We 
have here $4N_{c}+4-2N_{f}$ instead of $2N_{c}+2-N_{f}$
because of the symmetry $w\leftrightarrow -w$ implied by the 
orientifold. The rotated curve is now completely determined.

\section{$N=1$ SQCD}
\setcounter{equation}{0}

We study the $\mu\rightarrow\infty$ limit of our M fivebrane configuration
and  compare it with the known
results in $N=1$ supersymmetric gauge theory.  We have considered 
the rotation of the left NS5-brane, which corresponds to the asymptotic region
$t\sim v^{2N_{c}+2}$ before the rotation and to $w\rightarrow\infty, 
v\sim\mu^{-1} w$ and $t\sim\mu^{-2N_{c}-2}w^{2N_{c}+2}$ after the rotation.
We expect the relation $t\sim v^{2N_{c}+2}$ to hold
also in the $\mu\rightarrow \infty$ because the D4 branes still end
on the left NS5-brane in this limit.

In order to preserve this relation,  we should rescale $t$ by a factor 
$\mu^{2N_c+2}$
and introduce a new variable
\beq
\label{new}
\tilde{t} = \mu^{2N_{c}+2} t
\eeq
which will have the same dependence on $ v$ as $ t$ had before the rotation and
this corresponds to the shift of the origin in the direction of 
$(x^6, x^{10})$.
After putting  $y =t$ in (\ref{ah1}), the space-time is described by
\beq
\tilde{y}z=\mu^{2N_{c}+2}\Lambda_{N=2}^{4N_{c}+4-2N_{f}}\prod_{i=1}^{N_{f}}
(v^{2}-m_{i}^{2}),
\eeq
where $\tilde{y} =\mu^{2N_{c}+2} y$.
This equation describes a smooth surface in the limit
$\mu \rightarrow\infty$ provided the product
\beq
\mu^{2N_{c}+2}\Lambda^{4N_{c}+4-2N_{f}} 
\eeq
remains to be  finite. We define this
product as follows:
\beq
\Lambda_{N=1}^{2(3N_{c}+3-N_{f})} = \mu^{2N_{c}+2}
\Lambda_{N=2}^{4N_{c}+4-2N_{f}}
\eeq
which is nothing but the RG matching condition of the
four-dimensional field theory. Note that this spacetime and M fivebrane 
is under
the rotation groups $U(1)_{4,5}$ and $U(1)_{8,9}$ in appropriate 
way discussed before.
We want to see what are the deformations of the $N=2$ Coulomb branch after
the rotation and the limit $\mu\rightarrow\infty$. 

\subsection{Pure $Sp(N_{c})$ Theory}

Without matter, the curve (\ref{ah0}) describing the $N=2$ Coulomb
branch is given by:
\beq
t^{2} - C_{N_{c}+1}(v^{2},u_{k})t + \Lambda^{4N_{c}+4}_{N=2} = 0
\eeq
where $C_{N_{c}+1}(v^{2},u_{k}) = v^2B(v^2, u_{k})$.
This curve is completely degenerate at $(N_{c} +1)$ points on the Coulomb
branch.
At one these points, the curve has the following form
\beq
v^{2} = \Lambda_{N=2}^{4} t^{-1/(N_{c}+1)} + t^{1/(N_{c}+1)}.
\eeq
Thus its rotation is 
\bea
v^{2} &=& \mu^{2} \Lambda_{N=2}^{4} w^{-2} + \mu^{-2} w^{2} \\ \nonumber
    t &=& \mu^{-2N_{c}-2}w^{2N_{c}+2}. \
\eea
Now we rescale $t$ as $\tilde{t} = \mu^{2N_{c}+2} t$ and send 
$\mu$ to $\infty$ by keeping $\Lambda_{N=1}$ finite. It is easy to see that 
the curve becomes in this 
limit:
\bea
v^{2} &=& \Lambda_{N=1}^{6} \tilde{t}^{-1/(N_{c}+1)} \\ \nonumber
    w^{2} &=& \tilde{t}^{1/(N_{c}+1)}. \
\eea
where the RG matching condition is used.
 
\subsection{Introducing Massless Matter}

For the rotated configuration we use now the expressions for 
$v^{2}=P(w^{2})$ and
$t=Q(w^{2})$ given before in terms of new variables. 
We again introduce the rescaled $\tilde{t}$
which is then given by
\beq
\label{rot1}
\tilde{t} = w^{2(N_{c} + 1 - N_{f})} (w^{2} - w_{0}^{2})^{N_{f}}.
\eeq
For $ v$ and $ w$ we have the relation by remembering that the order parameters
$u_k$ are independent of $\mu$, are powers of $\Lambda_{N=2}$ and vanish
in the $\mu \rightarrow \infty$
\beq
\label{rot3}
\tilde{t} = \Lambda_{N=1}^{6N_{c}+6-2N_{f}}v^{2(N_{f}-N_{c}-1)}.
\eeq
When $\mu\rightarrow\infty$, the limit for (\ref{rot1}) and (\ref{rot2}) 
is given by the
behavior of $w_{0}\sim \mu\Lambda_{N=2}$. 
By using the relation:
\beq
\label{rel1}
\mu\Lambda_{N=2} = (\Lambda_{N=1}^{3N_{c}+3-N_{f}}
\mu^{N_{c}+1-N_{f}})^{\frac{1}
{2N_{c}+2-N_{f}}},
\eeq
we have three regions for $N_f$. Let us see how the curves look like: 

$\bullet \;\;\; N_{f} < N_{c} + 1$

(\ref{rel1}) tells us that $\mu\Lambda_{N=2}$ diverges and 
$w_{0}$ also diverges. Therefore the curve becomes infinite in the
$x^{6}$ direction. So there is no field theory in four dimensions. This is
just the same as saying that there is no supersymmetric vacua in the
$N=1$ theory.

$\bullet \;\;\; N_c+1 < N_{f} <  2(N_{c} + 1)$

In this case, from (\ref{rel1}) it is easy too see  that 
$\mu\Lambda_{N=2} = 0$ in the
limit $\mu\rightarrow\infty$ and (\ref{rot1}), (\ref{rot2}) and (\ref{rot3}) 
transform into:
\bea
\tilde{t} &=& w^{2(N_{c}+1)} \\ \nonumber
    v~w   &=& 0             \\ \nonumber  
 v^{2(N_{c}+1)}\tilde{t} &=& \Lambda^{6(N_{c}+1) - 2N_{f}} v^{2N_{f}}.\
\eea
As explained in \cite{hoo}, only the limit $\tilde{t}\ne 0, w=0$ is allowed,
so the interpretation of the previous equation is that the curve splits 
into two components in this limit: $C_{L} (\tilde{t} = w^{2(N_{c} + 1)}, 
v = 0)$
and $C_{R} (\tilde{t} = \Lambda_{N=1}^{6N_{c}+6-2N_{f}} v^{2N_{f}-2N_{c}-2}, 
w=0)$
where the component $C_{L}$ corresponds to the NS'5 brane which was
rotated and on the other hand, $C_{R}$ refers to the NS5 brane and the 
attached D4-branes.

${\bullet} \;\;\; N_{f} = N_{c} + 1$

In this case, the RG matching condition tells that 
$\mu\Lambda_{N=2}$ is equal to $\Lambda_{N=1}^{2}$. The equations 
(\ref{rot3}), (\ref{rot1}) and (\ref{rot2}) become:
\bea
\tilde{t} &=& \Lambda_{N=1}^{4(N_{c}+1)} \\ \nonumber
\tilde{t} &=& (w^{2}-w_{0}^{2})^{N_{c}+1} \\ \nonumber
vw        &=& 0. \
\eea   
The correct interpretation of these equations is that the curve also 
splits into
two components: $C_{L} (\tilde{t} = (w^{2} - w_{0}^{2})^{N_{c} + 1}, v = 0)$
and $C_{R} (\tilde{t} = \Lambda_{N=1}^{4(N_{c}+1)}, w = 0)$.

For $SU(N_c)$ group the cases $N_{f} = N_{c} + 1$ and $N_{f} > N_{c} + 1$
differed from each other because the first different non-baryonic 
branch roots went to
different limits and the second all non-baryonic branch roots have the
same limit. That was determined by the fact that $M$, the meson matrix, had 2 
different values for the diagonal entries. However in our case, for 
the $Sp(N_c)$ group,
there is only one kind of top-right diagonal entry, so we do not see 
those difference
appeared in $SU(N_c)$ gauge group and also do not have any baryonic branch.

\subsection{Massive Matter}

Before starting our discussion of introducing matter for our case, let 
us briefly examine the difference between the results of \cite{hoo}
and \cite{biksy,ss} for the case of massive matter when we consider
$SU(N_c)$ gauge group.
Actually we will
just compare the equation (5.28) of \cite{hoo}, and (4.6) and
(5.1) of \cite{ss}.
Notice that the second equation in (4.6) of \cite{ss} and the first 
one in (5.28) of
\cite{hoo} are the same:
\beq
v~w = (m_{f}^{N_{f}}\Lambda_{N=1}^{3N_{c}-N_{f}})^{1/N_{c}}.
\eeq
The first equation of (4.6) in \cite{ss} is not the same as the second one
of (5.28) in \cite{hoo}. Rather the equation of (5.28) looks
like the equation (5.1) of \cite{ss} because there we have the
dependence $t-w$. Recall that the vev for $M$ are given for equal
squark masses by:
\beq
m = m_{f}^{\frac{N_{f} - N_{c}}{N_{c}}}\Lambda_{N=1}^{
\frac{3N_{c}-N_{f}}{N_{c}}}.
\eeq
The relation between $\tilde{t}$ and $w$ can be rewritten as:
\beq
\label{equi1}
\tilde{t} w^{N_{f} - N_{c}} = (w - m)^{N_{f}}.
\eeq
 When we write it in terms of $t$ and $v$, this turns out:
\beq
\label{equi2}
\tilde{t} v^{N_{c}} = (-1)^{N_f} \Lambda_{N=1}^{3N_{c} - 
N_{f}} (v - m_{f})^{N_{f}}
\eeq
The relations (\ref{equi1}) and (\ref{equi2}) are just the equivalent of 
(4.6) and
(5.1) in \cite{ss}. In (\ref{equi2}) we have a supplementary 
power of $\Lambda$ as compared with \cite{ss}. This is due to the fact that
in \cite{hoo}, $t$ has a dimension of mass by its definition, but in 
\cite{ss} $t$ is dimensionless. The power of $\Lambda$ is just used 
to match the dimension of mass. We then find that the same curve 
can be written 
in terms of $ t-w$ and $t-v$. The two descriptions correspond to the $N=1$
duality \cite{se1}, between theories with gauge groups $SU(N_{c})$ and
$SU(N_{f}-N_{c})$, as we can see from the dependence of $t$ as a
function of $v$ and $w$ in 
(\ref{equi1}) and (\ref{equi2}). So the electric-magnetic duality 
can be observed also
from the set-up of \cite{hoo}. The connection between \cite{hoo} and
\cite{ss} may become clear only after we should introduce D6 branes in the
set-up of \cite{ss} which is a very interesting direction to pursue and 
investigate.

What happens when we consider our case? If all the quarks have equal mass
$m_{f}$, then the curve for $\mu\rightarrow\infty$ becomes:
\bea
\label{s0}
v^{2}w^{2} &=& (m_{f}^{N_{f}}\Lambda_{N=1}^{3(N_{c}+1)-N_{f}})^{2/(N_{c}+1)}
\\ \nonumber
\tilde{t}  &=& w^{2(N_{c}+1-N_{f})}\left(w^{2}-\left(\frac{ 2^{2N_c-N_f} 
\Lambda_{N=1}^{3(N_{c}
+1-N_{f})}}{m_{f}^{N_{c}+1-N_{f}}}\right)^{2/(N_{c}+1)}\right)^{N_{f}}. \
\eea
By the relation which connects the vev of $M$ and the masses of
quarks, we obtain:
\beq
\tilde{t} w^{2(N_{f}-N_{c}-1)} = (w^{2}-m^{2})^{N_{f}}.
\label{s1}
\eeq
When we  write it  in terms of $ t$ and $v$, this gives rise to:
\beq
\tilde{t} v^{2N_{c}+2} = (-1)^{N_f} \Lambda_{N=1}^{6(N_{c}+1)-2N_{f}} (v^{2}-
2^{\frac{2(2N_c-N_f)}{N_c+1}} {m_f}^{2})^{N_{f}}.
\label{s2}
\eeq
>From the relations (\ref{s1}) and (\ref{s2}) we again see the duality between 
the theories with gauge groups $Sp(N_{c})$ and $Sp(N_{f}-N_{c}-2)$.
The above relations are similar to the equations 
(2) and (6) of \cite{cs}. The equation between $ v$ and $w$ is just the same
as those in \cite{cs}. The electric-magnetic duality appears as an interchange
$v-w$, the curve describing the M-theory configuration being unique. Again,
in \cite{cs} $t$ is dimensionless while in our case $\tilde{t}$ has a
specific dimension. Therefore we see a power of $\Lambda$
in (\ref{s2}). So (\ref{s1}) and (\ref{s2}) give the electric-magnetic duality 
in our approach. It would be very interesting to introduce D6 branes
in \cite{cs} and to see the coincidence of the corresponding solution with
ours.

Now take the limit $m_{f}\rightarrow\infty$. After we  integrate
out the massive flavors and use the matching of the running coupling
constant 
\beq
\tilde{\Lambda}_{N=1}^{3(N_{c}+1)}=m_{f}^{N_{f}}
\Lambda_{N=1}^{3(N_{c}+1)-N_{f}}
\eeq 
to rewrite the equation (\ref{s0}) as
\bea
v^{2}w^{2} &=& \tilde{\Lambda}^{6}_{N=1} \\ \nonumber
\tilde{t} &=& w^{2(N_{c}+1-N_{f})} \left( w^{2} - 2^{\frac{2(2N_c-N_f)}{N_c+1}}
\frac{\tilde{\Lambda}^{6}_{N=1}}{m_f^{2}}\right)^{N_{f}}. \
\eea
If we keep $\tilde{\Lambda}_{N=1}$ finite while sending $m_{f}$ to $\infty$,
this again reduces to the pure Yang-Mills result.

\section{Conclusions}
\setcounter{equation}{0}

In the present work we considered the M theory description of the supersymmetry
breaking from $N=2$ to $N=1$ for the case
of symplectic gauge group $Sp(N_c)$ obtaining many aspects of 
the strong coupling
phenomena. In this case  
no baryon can be constructed so the only Higgs branch is the
non-baryonic branch.
In field theory approach, starting with a $N=2$ supersymmetric gauge theory and
giving mass to the adjoint chiral multiplet, the extremum of the superpotential
gave us a unique solution for the expectation value for the meson matrix
$M$ in which $\mbox{Tr}M=0$ as opossed to the
$SU(N_{c})$ case where $\mbox{M}\ne 0$. 

In the M theory fivebrane approach, we discussed first 
the unrotated configuration
which corresponds to an $N=2$ theory, with or without D6 branes. For
the case without D6 branes, M theory fivebrane configuration is a
single fivebrane with the world volume ${\bf R^{1,3}} \times \Sigma$ where
$\Sigma$ is the Seiberg-Witten curve of the gauge group $Sp(N_c)$. 
By introducing D6 branes, we have 
considered the complex structure of the corresponding Taub-NUT space 
which is the same one as those of the ALE space of
$A_{2n-1}$-type and resolved the $A_{2n-1}$ singularity.

One of the most important aspects of the $Sp(N_{c})$ gauge theory was the O4
orientifold which is parallel to D4 branes . Its antisymmetric projection
which eliminates some degrees of freedom was essential in matching the
dimension of the Higgs moduli space in IIA  brane approach with the one
of the field theory. This observation was used not only in type 
IIA picture when
counting the number of D4 branes suspended between the D6 branes but also
in the M theory picture when counting the multiplicities of the
rational curves. It is known \cite{hov,vafa}
that A type singularity by imposing the ${\bf Z_2}$ symmetry, due to
the orientifolding, leads to D type singularity.
We expect that our M theory fivebrane argument starting from D type singularity
can go similarly and will see how the interrelation between two 
types singularities plays 
a role.

For rotated branes we used the coordinates 
$v=x^{4}+ix^{5}$ and $w=x^{8}+ix^{9}$, the position of
the D4 branes in the $w$ direction being identified with the eigenvalues of the
meson matrix $M$. We found at most one eigenvalue $w_{0}$ for the asymptotic
position of the D4 branes which is consistent with the field 
theory result where
only one eigenvalue for $M$ was found. We connected $w_{0}$ with the 
unique eigenvalue of $M$. In section 5 we have obtained the forms for the
rotated curves, for pure gauge group and for massive and massless
matter.

In all of our discussions as well as in many exciting works which appeared
recently, many results obtained in field theory were rederived
in M theory which makes M theory approach an extraordinary laboratory
to derive results which were very difficult to obtain only by 
pure field theory methods.
Now we have a clearer view over the strongly coupled phenomena of
supersymmetric theories. But we still need to obtain new information
like how to introduce flavors in the spinor representation and how to
obtain dualities for $N=1$ theories with gauge groups like as 
$SO\times SO$ or
$SO\times SU$, for which M theory approach did not give yet any 
supplementary information compared with type IIA approach. We hope that
this information will be reached in the near future.

\end{document}